\documentclass[a4paper,12pt,nohyper]{JHEP3}

\newcommand{\bea}{\begin{eqnarray}}
\newcommand{\beal}[1]{\begin{eqnarray}\label{#1}}
\newcommand{\eea}{\end{eqnarray}} 
\newcommand{\be}{\begin{equation}} 
\newcommand{\bel}[1]{\begin{equation}\label{#1}}
\newcommand{\ee}{\end{equation}} 
\newcommand{\rf}[1]{(\ref{#1})}

\newcommand{\bit}{\begin{itemize}}
\newcommand{\eit}{\end{itemize}}
\newcommand{\ben}{\begin{enumerate}}
\newcommand{\een}{\end{enumerate}}

\def\half{\frac{1}{2}}

\def\quart{\frac{1}{4}}

\def\alp{\leavevmode\ifmmode {\alpha^\prime} \else ${\alpha^\prime}$ \fi}

\title{Some half-BPS solutions of M-theory}

\preprint{}

\author{Micha\l\ Spali\'nski\\
So\l tan Institute for Nuclear Studies\\
ul. Ho\.za 69, 
00-681 Warszawa, Polska.
}

\abstract{
It was recently shown that half BPS-solutions of M-theory can be expressed
in terms of a single function 
satisfying the 3d continuum Toda equation. In this note half-BPS solutions
corresponding to separable 
solutions of the Toda equation are examined. 
}


\begin{document}

\section{Introduction}

It was recently shown by Lin, Lunin and Maldacena (LLM) that half
BPS-solutions of M-theory can be expressed in terms of a single function
$D$ satisfying the 3d continuum Toda equation\cite{Lin:2004nb}.  Solutions
to this equation are not easily found. There is however a simple class of
solutions which turns out to lead to physically interesting geometries,
namely separable solutions which can locally be expressed in terms of a
quadratic form in a single variable and solutions of the Liouville
equation.

Solutions of 11-dimensional supergravity are of great interest and have
been discussed extensively in the literature (see for example
\cite{Gauntlett:1997cv}-\cite{Gauntlett:2004zh} and references therein).
The approach described in \cite{Lin:2004nb} provides a way to study them in
a unified framework. The moduli space of half-BPS solutions is of great
interest from the point of view of the AdS/CFT
correspondence\cite{Maldacena:1997re}. Already some very interesting
physical questions (such as topology change \cite{Horava:2005pv}) have been
addressed using this approach (in particular the relation to the phase
space of free fermions\cite{Berenstein:2004kk}).

The LLM description applies to both regular and singular solutions. 
For a regular solution the function $D$  has to satisfy specific boundary 
conditions. Generic solutions for $D$ will not satisfy these boundary
conditions and will give rise to singular geometries, which may be
interpreted as a sign that important degrees of freedom have been
neglected. 

It turns out that among the geometries corresponding to separable solutions
of the Toda equation there is only one non-singular example -- the
Maldacena-Nunez solution\cite{Maldacena:2000mw}, which describes the
near-horizon limit of M5-branes wrapping holomorphic curves in Calabi-Yau
threefolds.  All the other possibilities which stem from separable
solutions of the Toda equation turn out to have singularities if standard
signature is imposed.

Two special cases are discussed in this note. One is the Alishahiha-Oz
solution\cite{Alishahiha:1999ds}, which describes a system of intersecting
M5-branes.  This solution is a warped product of $AdS_5$ and a
6-dimensional space and as such is dual to a superconformal field theory in
four dimensions. The other example discussed below is a warped product with
an $AdS_2$ factor, which is expected to be dual to a model of
superconformal quantum mechanics. This geometry describes the near-horizon
limit of a system of intersecting M2-branes\cite{Cvetic:2000cj}. Warped
products with $AdS$ factors are of great interest from various points of
view, including the AdS/CFT correspondence as well as flux
compactifications to four dimensions.

This note begins by citing the relevant formulas from \cite{Lin:2004nb},
after which the separable solutions of the Toda equations are
described. Rather than analyzing all the cases systematically, only 
some general features are given and a few 
special cases are discussed, possibly leaving a more complete presentation
for the future.

\section{LLM solutions}

The general solution of the supergravity equations of motion described by
LLM has the form 
\beal{metric1} 
ds_{11}^2 &=& -{4e^{2\lambda} (1+ y^2e^{-6\lambda})}(dt+V_idx^i)^2 + 
4e^{2\lambda} d\Omega_5^2+y^2e^{-4\lambda}d{\tilde\Omega}_2^2 +\nonumber \\
&+& \frac{e^{-4\lambda}}{1+y^2e^{-6\lambda}}[dy^2 + e^{D} (dx_1^2 + dx_2^2)]\\
e^{-6\lambda} &=& {\partial_y  D \over y(1 -  y \partial_y D) } \\
V_i &=& \half \epsilon_{ij} \partial_j D \ ,
\eea
where $i,j=1,2$. This, supplemented by the fluxes given in
\cite{Lin:2004nb}, 
provides an M-theory background preserving 16 of the original 32
supersymmetries. 

The function $D$ which determines the solution obeys the equation
\bel{toda}
(\partial_{x_1}^2 + \partial_{x_2}^2)D + \partial_y^2 e^{D} =0 \ .
\ee
This is the 3-dimensional continuous version of the Toda equation. To
obtain regular supergravity solutions one has to impose specific boundary
conditions on 
$D$ so that potential singularities inherent in \rf{metric1} do not
occur. These conditions can be found in \cite{Lin:2004nb}.  

Note that the form of the ansatz is preserved under $y$-independent
conformal transformations of the $x_1-x_2$ plane provided $D$ is shifted  
appropriately: 
\be
x_1 + i x_2 \to g(x_1 + i x_2), \quad D \to D - \log |\partial g|^2 \ .
\ee

The solutions examined above can be Wick-rotated (as discussed by LLM) 
to yield solutions of 11-dimensional supergravity which contain an $AdS_5$
factor and a compact 6-dimensional manifold. These solutions can be
interpreted as dual to conformal field theories in four dimensions. 
Specifically, one can get solutions of the form of an $AdS_5$ warped
product by performing the analytic continuation 
\bel{wick1a}
\psi \rightarrow \tau \quad \alpha \rightarrow i \rho\nonumber\ .
\ee
This maps 
\be
\cos^2 \alpha d\psi^2 + d \alpha^2 + \sin^2 \alpha d\Omega_3^2 
\rightarrow
 - ( - \cosh^2 \rho d\tau^2 + d \rho^2 + \sinh^2 \rho d \Omega_3^2 )
\ee
i.e. 
\be
d\Omega_5^2  \rightarrow  -ds_{AdS_5}^2 \ .
\ee
In addition one has to take\cite{Lin:2004nb} 
\bel{wick1b}
\lambda  = \tilde \lambda + i { \pi \over 2} \ ,
\ee
with the remaining coordinates unchanged. 
For real $\tilde \lambda$ one finds a metric with the correct
signature\footnote{Since $t$ is now a spacelike coordinate, it is 
denoted by $\chi$ below to avoid confusion.}. 
This way one arrives at 
\bel{metric2}
ds_{11}^2 = e^{2\tilde \lambda}\left( 4
ds_{AdS_5}^2+y^2e^{-6{\tilde\lambda}} d{\tilde\Omega}_2^2+ds_4^2\right)\ ,
\ee
where 
\be
ds_4^2= {4 ( 1 - y^2e^{-6 \tilde \lambda})} (d\chi+V_idx^i)^2 +
\frac{e^{-6\tilde\lambda} }{1 - y^2e^{-6\tilde \lambda}} [ dy^2 + e^{D}
(dx_1^2 + dx_2^2) ]  
\ee
and
\be
e^{-6\tilde \lambda}= - {\partial_y D \over y(1 -  y \partial_y D) } \ ,
\ee
where $D$ satisfies the Toda equation \rf{toda}. Note that 
due to the analytic continuation 
regular
solutions in this case have to satisfy different boundary
conditions than those for \rf{metric1}. The form \rf{metric2} characterizes
all M-theory compactifications to $AdS_5$ which preserve $N=2$
supersymmetry in four dimensions. 

Instead of the Wick rotation \rf{wick1a}, \rf{wick1b} one can instead do
\be
y\rightarrow i y\ , \quad x_k \rightarrow i x_k \ ,
\ee
which leads to metrics of the form of a warped product of $AdS_2$ with a
9-dimensional manifold:
\bel{metric3}
ds_{11}^2=e^{2\tilde\lambda}\left( 4 
ds_{S^5}^2+y^2e^{-6{\tilde\lambda}} ds^2_{AdS_2}- ds_4^2\right) \ ,
\ee
where $ds_4^2$ and $e^{-6{\tilde \lambda}}$ are as above. The form
\rf{metric3} characterizes 
all M-theory compactifications to $AdS_2$ which preserve $N=2$
supersymmetry in four dimensions.

\section{Separable solutions of the Toda equation}

Separable solutions of \rf{toda} are of the form\footnote{In a different
  context such solutions were discussed in
  \cite{Calderbank:1999ad,Ketov:2000ni}.}:  
\be
D(x_1,x_2,y) = F(x_1, x_2) + G(y) \ .
\ee
Using this in the Toda equation \rf{toda} one finds that 
\be
G(y) = \log (\alpha y^2 + \beta y + \gamma) \ ,
\ee
while $F$ has to satisfy the Liouville equation:
\be
(\partial^2_{x_1} + \partial^2_{x_2}) F + 2 \alpha e^F = 0 \ .
\ee
Here $\alpha$, $\beta$ and $\gamma$ are constants. 

In terms of $\xi\equiv x_1+i x_2$ one has the well known general
solution of the Liouvile equation\cite{jost}: 
\bel{lsol}
e^F = \frac{4 |f'(\xi)|^2}{(1+\alpha |f(\xi)|^2)^2} \ ,
\ee
where $f$ is a holomorphic function. 

Thus the function $D$ is given by
\bel{dsol}
D(x_1,x_2,y) = \log \frac{4 Q(y) |f'(\xi)|^2}{(1+\alpha |f(\xi)|^2)^2}
\ee
with 
\be
Q(y) = \alpha y^2 + \beta y + \gamma\ .
\ee
The parameters $\alpha, \beta, \gamma$ parameterize the possible
solutions. It is convenient to discuss separately the following cases: 
\begin{enumerate}
\item $\alpha=0$, i.e. $Q=\beta y + \gamma$;
\item $\alpha<0$ and $Q=-|\alpha| (y-y_1) (y-y_2)$ for real $y_1, y_2$;
\item $\alpha>0$ and $Q= \alpha (y-y_1) (y-y_2)$ for real $y_1, y_2$; 
\item $\alpha>0$  and $Q= \alpha |y - u|^2$ for complex $u$. 
\end{enumerate} 
and it will be assumed that $y_1<y_2$. 
In the first case $\beta$ can be scaled away by redefining the function
$f$ in \rf{dsol}. In the remaining cases $\alpha$ can be scaled away in the
same way once its sign is fixed. 

Since the Toda equation is invariant under  
\be
\xi \longrightarrow g(\xi)
\ee
for any holomorphic function $g$, one can locally choose a convenient
canonical form for the function $G$. 
In view of this, one can take the following solutions for $D$:
\bea
D_0(x_1,x_2,y) &=& \log (y+\gamma) \label{d0}\ ,\\
D_1(x_1,x_2,y) &=& \log \frac{(y-y_1)(y_2-y)}{x_2^2} \label{d1}\ ,\\
D_2(x_1,x_2,y) &=& \log \frac{4 (y-y_1)(y-y_2)}{(1+(x_1^2+x_2^2)^2)^2} \ ,
\label{d2}\\ 
D_3(x_1,x_2,y) &=& \log \frac{4 |y - u|^2 }{(1+(x_1^2+x_2^2)^2)^2}\ . \label{d3}
\eea

\section{M-theory solutions}

Each of the solutions \rf{d0}--\rf{d3} for $D$ leads to an $N=2$
supersymmetric solution of the 
supergravity equations of motion. However to interpret the resulting
metrics in physical terms one has to properly define the ranges of
coordinates so that the signature is correct. 
All but one of the metrics arising
from \rf{d0}--\rf{d3} are singular at some points. Such 
singularities are assumed to indicate that some degrees of freedom which
generically decouple become light\cite{Alishahiha:1999ds,Oz:1999qd} and
have to be accounted for if a non-singular description is to ensue. 

A case which is regular was already pointed out in \cite{Lin:2004nb},
where the 
following solution of the Toda equation is discussed: 
\bel{maln}
e^{D} = \frac{1}{x_2^2} (1 - 4y^2) \ .
\ee
This leads to the regular geometry found earlier by Maldacena and 
Nunez\cite{Maldacena:2000mw}. The form \rf{maln} is clearly separable and
is in fact 
of the form $D_1$ given in \rf{d1}. Substituting the solution \rf{d1} in
\rf{metric2} yields (note that $y_1<y<y_2$):
\newcommand{\ysum}{y_1+y_2-2y}
\newcommand{\yprod}{y_1 y_2-y^2}
\newcommand{\yysum}{\frac{y}{\ysum}}
\newcommand{\ysumy}{\frac{\ysum}{y}}
\bea 
ds^2 &=& (\yprod)^{1/3} \left(4(\yysum)^{1/3} ds_{AdS_5}^2 +
\frac{y^{4/3} (\ysum)^{2/3}}{\yprod}\ d\Omega_2^2 +\nonumber\right. \\
&+&\left. 4 (\yysum)^{1/3} \frac{(y-y_1)(y-y_2)}{\yprod}
(d\chi-\frac{dx_2}{x_2})^2 +\nonumber\right. \\
&+& \left. 
(\ysumy)^{2/3} \left(\frac{dy^2}{(y-y_1)(y-y_2)} -
\frac{dx_1^2 + dx_2^2}{x_2^2}\right) \right) \ .
\eea
This  metric is  singular
if the roots $y_1, y_2$ are arbitrary. 
However if one sets $y_1=-s, y_2=s$ (for
some real $s$), then the metric is regular (once $\chi$ is identified 
with period $2\pi$):
\beal{gmn}
ds^2 &=& 2^{2/3} (s^2 + y^2)^{1/3} \left(2 ds^2_{AdS_5} + \frac{y^2}{s^2+y^2}
d\Omega_2^2 \right. \nonumber\\
&+& \left. 2\frac{s^2-y^2}{s^2+y^2} (d\chi-\frac{dx_1}{x_2})^2 +
\frac{dy^2}{s^2-y^2} + \frac{dx_1^2 + dx_2^2}{x_2^2}\right)\ .
\eea
By rescaling $y$ one can reduce the $s$-dependence to an overall factor. 
Substituting $y=s\cos\theta$ one finds the metric 
\beal{mnmetr}
ds^2 &=& (2s)^{2/3} \Delta^{1/3} \left(2 ds^2_{AdS_5} + \Delta^{-1}
\cos^2\theta 
d\Omega_2^2 + \right. \nonumber \\
&+& \left. 2\Delta^{-1} \sin^2\theta (d\chi-\frac{dx_1}{x_2})^2 +
d\theta^2 + \frac{dx_1^2 + dx_2^2}{x_2^2} \right)\ ,
\eea
where $\Delta=1+\cos^2\theta$, which is up to an overall factor 
the same as the metric appearing in \cite{Maldacena:2000mw}. The metric
\rf{mnmetr} has the form of an $AdS_5$ fibration, and so is 
expected to be the supergravity dual of a superconformal field theory in
four dimensions defined on the boundary of $AdS_5$. It may be
interesting to study the singular deformations of \rf{mnmetr} described by
\rf{gmn}.

Another previously known case is a singular solution first discussed in 
\cite{Alishahiha:1999ds} as the description of the 
near-horizon limit of a system of intersecting M5-branes. 
In the present context it arises by using solution $D_0$, eq. \rf{d1}, in
the metric \rf{metric2}.  To obtain the correct signature one needs 
$\gamma>0$
and $-\gamma < y < 0$. The metric can be written as 
\be
ds^2 = - \gamma^{1/3} y^{1/3} \left(4 ds^2_{AdS_5} -\frac{y}{\gamma}
d\Omega_2^2 + 
\frac{y+\gamma}{\gamma} d\chi^2 - \frac{dy^2}{y (y+\gamma)} - \frac{dx_1^2 +
  dx_2^2}{y}\right)\ . 
\ee
The change of variable $y=-\gamma\sin^2\alpha$ results in 
\be
ds^2 = 4\gamma^{2/3} \sin^{2/3}\alpha \left(ds^2_{AdS_5} + \quart
\sin^2\alpha\ d\Omega_2^2 + d\alpha^2 +\cos^2\alpha\ d\chi^2 +
\frac{dx_1^2+dx_2^2}{\gamma\sin^2\alpha} \right)\ ,
\ee
which is of the form given in \cite{Alishahiha:1999ds}. The
singularity at $\alpha=0$ was interpreted there as being due to M2-branes
ending on the M5-branes. 

A somewhat similar example 
arises from using solution $D_0$ \rf{d0} in the
metric \rf{metric3}. 
To have the 
correct signature one needs $\gamma < 0$
and $y >|\gamma|$. The metric reads
\bea
ds^2 &=& \frac{y^{4/3}}{\gamma^{2/3}} \left(ds^2_{AdS_2} + 4|\gamma|
d\Omega_5^2    
+ \frac{|\gamma|}{y^2} (dx_1^2 + dx_2^2) + \right. \nonumber \\
&+& \left.4 \frac{y-|\gamma|}{y} d\chi^2 + \frac{|\gamma|}{y^2(y-|\gamma|)}
dy^2\right)   \ .
\eea
Setting $\gamma=-1$ and $y=1/\sin^2\alpha$ leads to 
\be
ds^2 = \sin^{-8/3}\alpha \left(ds^2_{AdS_2} + d\alpha^2 + 
\cos^2\alpha\ d\chi^2 
+ \sin^2\alpha\ d\Omega_5^2  + \sin^4\alpha\ (dx_1^2 + dx_2^2)\right)\ ,
\ee
which is the form found in \cite{Cvetic:2000cj} for the near-horizon
geometry of a 
system of semilocalised intersecting M2-branes. 
As this is an example of an $AdS_2$ fibration it can be expected to be dual
to a superconformal quantum mechanics.

\section{Conclusions}

Although very simple, separable solutions of the Toda equation lead to a
large family of half-BPS solutions of M-theory which include at least some
physically interesting cases. 
All but one of the solutions arising this way are singular. It would be
interesting to establish whether these 
geometries have interpretations in terms of branes. 
It could also be of interest to explore these
solutions more closely, in particular, to understand the origin and
interpretation of their
singularities. 
Another natural question is whether there is a geometric interpretation of 
parameters appearing in the quadratic form $Q$. 

It is perhaps disappointing that this
class of supergravity solutions does not 
include any regular cases beyond the well-known Maldacena-Nunez
solution. To find new regular cases one has to
understand more general (in particular non-separable) solutions of the Toda
equation. One example of a 
non-separable solution is in fact determined by the functions $D_3$ in 
eq. \rf{d3} if one allows the parameter $u$ appearing there to depend
holomorphically on
$\xi\equiv x_1+i x_2$. 
Such solutions were introduced and studied by Calderbank and
Tod\cite{Calderbank:1999ad}. While non-separable, these solutions also do
not lead to regular metrics. It appears that for this purpose the
construction described by Ward\cite{ward} (which is discussed  
by LLM), may be more promising.

\end{document}